# Signatures of moiré intralayer biexcitons and exciton-phason coupling in $WSe_2/WS_2$ heterostructures

Ranju Dalal[1], Harsimran Singh[1], Rwik Dutta[2], Hariharan Swaminathan[1], Kenji Watanabe[3], Takashi Taniguchi[4], Mit H Naik[2*], Manish Jain[1*], Akshay Singh[1*]

[1]Department of Physics, Indian Institute of Science, Bengaluru, Karnataka 560012, India

[2]Department of Physics and Center for Complex Quantum Systems, University of Texas at Austin, Austin, Texas 78712, USA

[3]Research Center for Functional Materials, National Institute for Materials Science, Ibaraki 3050044, Japan

[4]International Center for Materials Nanoarchitectonics, National Institute for Materials Science, Ibaraki 3050044 Japan

*Corresponding author: mit.naik@austin.utexas.edu, mjain@iisc.ac.in, aksy@iisc.ac.in

**Abstract:**

Interactions among electronic and lattice degrees-of-freedom are foundational to various phases in condensed-matter physics, yet the dynamic interplay between excitonic and phononic quasiparticles represents an equivalent, underexplored frontier. Moiré superlattices provide an ideal platform for realizing these interactions by offering localized intralayer excitons (IALX) and ultralow-energy collective lattice modes, such as phasons. Here, by optically suppressing ultrafast charge-transfer (CT) to interlayer excitons in $WSe_2/WS_2$ heterostructures, we uncover dynamics of moiré IALX revealing long lifetimes ($\tau > 1000$ ps) arising from localized Wannier and in-plane CT nature. We then observe moiré intralayer intervalley biexcitons with binding energy $\sim 16$ meV, with long lifetimes due to moiré confinement. Furthermore, we find time-domain signatures of strong coupling between moiré-IALX and $\sim 10$ μeV phasons, evidenced as twist-angle-dependent GHz oscillations in IALX dynamics. Our findings establish moiré superlattices as interacting hybrid quantum systems and for engineering non-equilibrium phenomena, as well as for GHz-scale optoelectronics.

**Introduction:**

Collective phenomena in two-dimensional (2D) semiconducting transition metal dichalcogenides (TMDs) are induced by the interplay of emergent quasiparticles, primarily excitons and phonons[1]. The strong binding of excitons fosters a hierarchy of many-body complexes, including trions and biexcitons[2–4], that dominate optical response, while phonons govern key processes such as exciton-diffusion and electronic and thermal properties[5–7]. Interactions among quasiparticles are foundational for hosting emergent quantum phenomena, from unconventional superconductivity[8] to topological phases and polaronic crystals[9], and underpin applications in optoelectronics and valleytronics[10].

These interactions are fundamentally governed by electronic and phononic band structures, which can be modified by creating twisted 2D heterostructures (moiré superlattices)[11,12]. The superlattice introduces a long-wavelength moiré potential driven by structural reconstruction[13], which gives rise to localized quantum states with tuneable properties[14–16], as well as unique, low-frequency moiré phonons[17–19]. Phasons, the ultrasoft shear phonon modes (~ 10 µeV) inherent in moiré superlattices, result from relative translation between layers of heterostructures. This collective motion continuously alters interlayer atomic registries across entire superlattice[20], and can induce changes in electronic phases[21]. However, probing lattice dynamics and energetics of phasons is hindered by their extremely low frequencies, which lie beyond the resolution of conventional spectroscopic techniques.

Furthermore, while lower-energy interlayer excitons (IELX) in moiré systems are increasingly understood[16,22], properties of their fundamental precursors, intralayer excitons (IALX), remain largely unexplored. Studying IALX in heterostructures with type-II band alignment is challenging because the rapid interlayer charge transfer (CT) process preferentially populates IELX, thereby obscuring the dynamics of IALX along with their higher-order complexes. For example, while moiré interlayer biexcitons and trions have been observed[16,23,24], there are no corresponding reports for moiré intralayer biexcitons. Moiré IALX are especially interesting due to distinct exciton resonances of both localized Wannier-type and in-plane CT-type character[25]. Therefore, investigating dynamics of these IALX states, their complexes, and coupling to phason-like collective modes is essential for understanding emergent physics of moiré systems.

Here, we use WSe$_2$/WS$_2$ moiré superlattice as a model system to study IALX and their interaction with phason modes by combining pump-probe spectroscopy with ab initio GW Bethe-Salpeter equation (BSE) calculations for optical spectra and real-space distribution of moiré IALX, and force-field-based phonon/phason calculations. High excitation repetition-rate suppresses the dominant interlayer CT channel, thus enabling study of IALX dynamics. We observe long-lived moiré Wannier excitons (relaxation-time, $\tau \sim$ 1000 ps), with CT excitons having even longer dynamics ($\tau \gg$ 1000 ps), compared to few ps in monolayer WSe$_2$. Importantly, we provide evidence for moiré intralayer intervalley biexciton, a correlated four-particle state with binding energy $\sim$ 16 meV, whose enhanced stability ($\tau \sim$ 1000 ps) is a direct consequence of moiré confinement. Remarkably, we observe polarization-independent, twist-angle-dependent GHz oscillations in IALX dynamics, which we attribute to the time-domain measurement of exciton-phason coupling. Observing complex excitonic states, along with collective ultrasoft lattice modes, reveals intricate relationship between excitonic and lattice phenomena in moiré systems.

**Results and Discussion:**

**Moiré intralayer excitons (IALX) and their dynamics**

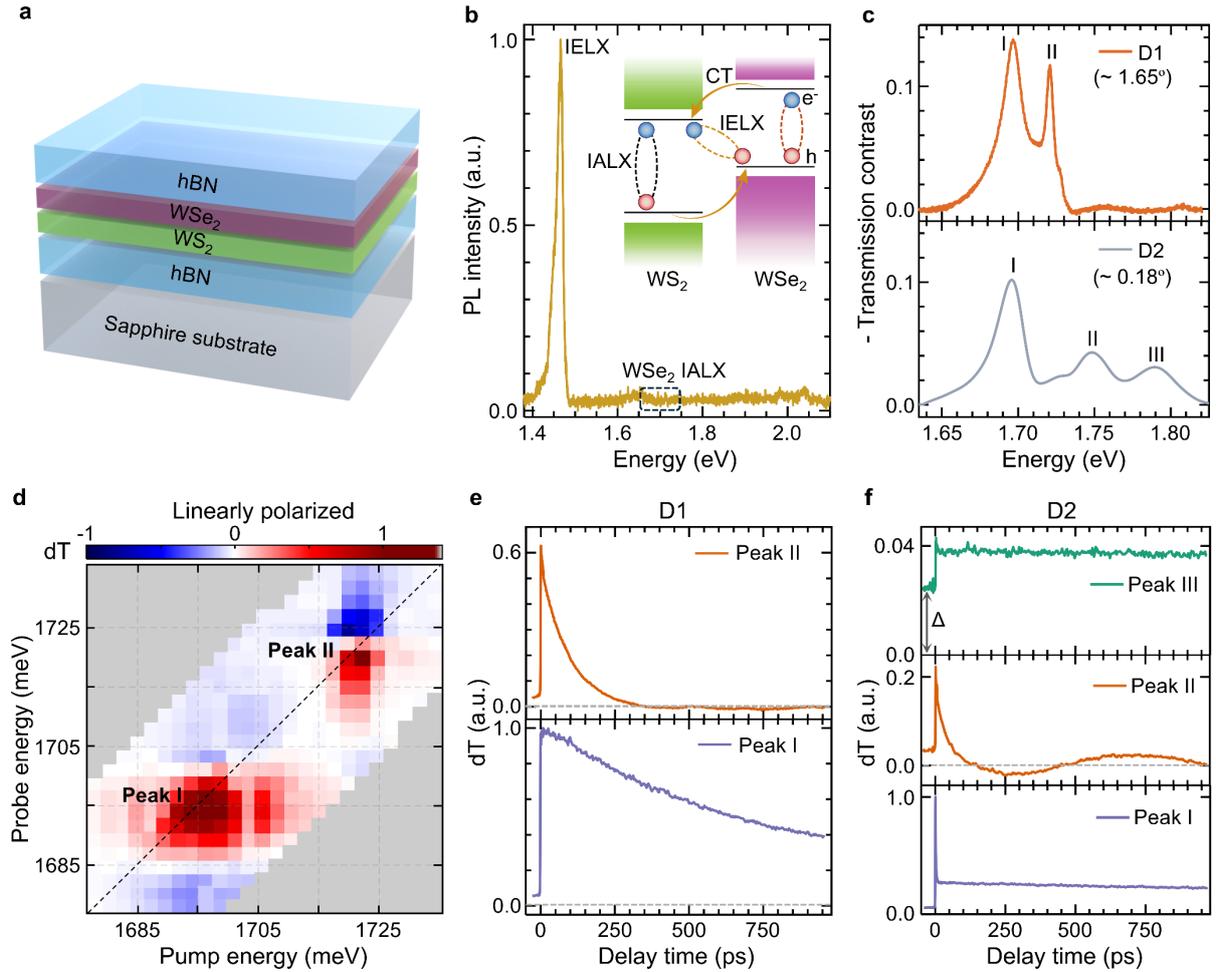

**Fig. 1: Moiré intralayer excitons (IALX) and their relaxation dynamics in $WSe_2/WS_2$. a,** Schematic shows hBN encapsulated $WSe_2/WS_2$ moiré heterostructure on sapphire substrate. **b,** Photoluminescence (PL) of R-stacked device, D1, shows strong interlayer exciton (IELX) emission centred at ~1.45 eV and highly quenched IALX due to layer-dependent charge transfer (CT) (schematic in inset). **c,** Steady-state transmission contrast from devices D1 (~ 1.65°) and D2 (~ 0.18°) shows two and three moiré IALX states, respectively. **d,** Linearly polarized (LP) two-colour pump-probe (TCPP) map of D1 at 2.5 ps delay time, with strong photobleached signals near resonances of peak-I and peak-II, along with spectral shift and a pump-induced absorption signal at probe energies lower than peak-I. **e, f,** Degenerate transient differential transmission (dT) of peak-I and peak-II of D1 **(e)** and peak I, II, and III of D2 **(f)** show distinct relaxation dynamics of different moiré IALX states, with small negative delay signals (Δ).

We prepare $WSe_2/WS_2$ heterostructures, encapsulated with hexagonal boron nitride (hBN), on sapphire substrates (Fig. 1a, Methods). Measurements are performed at 4 K in transmission geometry to remove ambiguity in the sign of differential-transmission (dT). Second-harmonic generation (SHG) measurements (Methods) confirm R-type stacking (Supplementary Section 1), while twist angles are measured using optical images for devices D1 (~ 1.65º) and D2 (~ 0.18º) (Supplementary Section 2). Photoluminescence (PL) spectrum (Methods) of D1 (Fig. 1b) is dominated by IELX emission at ~ 1.45 eV. Upon photoexcitation, due to type-II band alignment, efficient CT between layers leads to formation of IELX and highly quenched IALX PL (Fig. 1b, inset). To investigate moiré IALX of $WSe_2$, we first measure steady-state dT (Methods) of D1 and D2 (PL shown in Supplementary Section 3), revealing two and three IALX resonances near $WSe_2$ A-exciton energy[12] (Fig. 1c), consistent with their twist angle of 1º-2º and ~ 0º, respectively. These multiple IALX are clearly differentiated from a reference $WSe_2$ monolayer (Supplementary Section 4), and originate from structural reconstruction of $WSe_2/WS_2$ moiré superlattice[25,26], which leads to strain redistributions in individual layers.

We measure dynamics of these moiré IALX using two-colour pump-probe (TCPP) spectroscopy (Methods, Supplementary Section 5). Fig. 1d shows a 2D transient dT map for D1, recorded at a delay time of 2.5 ps, using linearly polarized (LP) pulses. A positive signal indicates photobleaching (PB), while negative signal signifies photo-induced absorption (PA). Grey region of the map is inaccessible due to tuning range limitations of TCPP. The map is dominated by two strong PB features from moiré IALX resonances I and II, accompanied by peak broadening and energy shifts analyzed in subsequent sections. A negative peak is observed when pumped at peak-I and probed at lower energies, the origin of which we discuss later.

Peak-I persists for a drastically longer time in D1 ($\tau$ ~ 1000 ps), c.f. A-exciton in $WSe_2$ monolayer[27]. Peak-II decays on a faster timescale of ~ 100 ps (Fig. 1e), with longer-lived oscillations (discussed later). Peak-I and peak-II dynamics are qualitatively similar in D2, which features an additional long-lived state, peak-III (Fig. 1f), with a weak signal consistent with its lower oscillator strength (Fig. 1c).

The observed dynamics of moiré IALX are effectively decoupled from IELX pathway by performing TCPP at 80 MHz. At this high excitation repetition-rate (~ 12.5 ns), we propose that long-lived (> 500 ns) IELX saturate and effectively block the ultrafast interlayer CT channel. We support this hypothesis by measuring repetition-rate dependent time-resolved PL (TRPL) (Supplementary Section 6). At a lower repetition-rate, CT is partially allowed, whose variable efficiency is a source of ambiguities in reported IELX dynamics[28–30]. Earlier transient-absorption measurements on similar systems were performed at ~ 100 kHz excitation rate,[28,31] where CT dominates IALX dynamics (Supplementary Section 7). The small negative delay signals (Δ, Fig. 1e, f) are attributed to quasi-static hole population in $WSe_2$, sustained by long-lived IELX.

**Nature of moiré IALX states**

Dynamics of moiré IALX can be understood by considering their excitonic wavefunctions within reconstructed $WSe_2/WS_2$ superlattice, as revealed by first-principles GW-BSE calculations using our recently developed pristine unit-cell matrix projection method[25]. The strain modulation in $WSe_2$ layer gives rise to a moiré potential that produces flat electronic bands and real-space modulation of valence and conduction states derived from *K*-valley of unit-cell Brillouin zone[26]. This modulation significantly alters the optical response, resulting in three distinct moiré exciton peaks with unique spatial characteristics[32] (Fig. 2 and Supplementary section 8). Peak-I corresponds to a Wannier-type exciton localized at AA stacking site[33] (top panel of Fig. 2), which acts as the global minimum of moiré potential. This strong spatial confinement can suppress non-radiative recombination by minimizing coupling to scattering channels, thereby enabling its extended lifetime. Additionally, to elucidate the evolution of exciton wavefunction with twist angle, we used BSE calculations with continuum-model-based-single-particle envelope functions (Supplementary Section 9), which show that qualitative character of the peaks remains largely unchanged for small deviations from 0° twist. However, oscillator strength of peak-I increases with increasing twist angle and the exciton becomes more delocalized (Supplementary Section 9), enhancing the radiative decay rate.

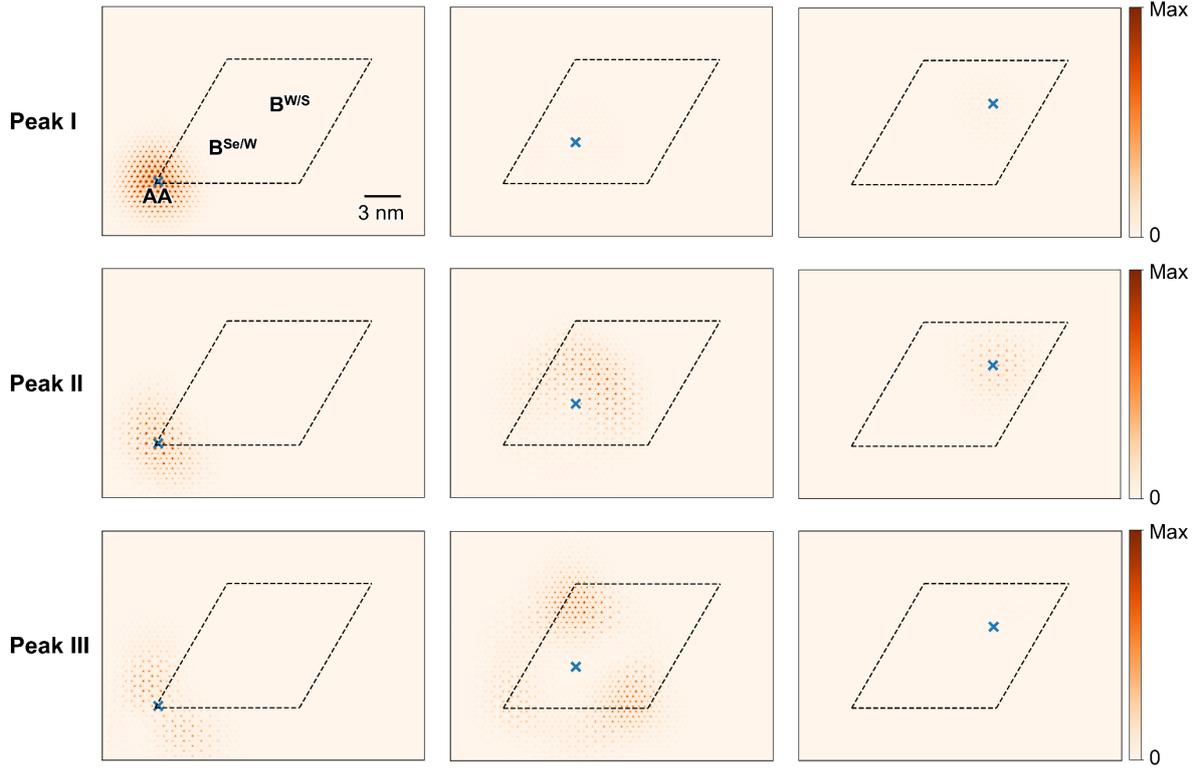

**Fig. 2: Real space distribution of moiré IALX.** Distribution of electron density as a function of fixed hole position (blue cross) in moiré unit cell for distinct IALX resonances shows a Wannier exciton for peak-I (top panel), CT exciton for peak-III (bottom panel) and a hybrid character for peak-II (middle panel).

Peak-III has an intralayer CT-exciton character with highest density of photoexcited hole at $B^{Se/W}$ site and highest photoexcited electron density at adjacent AA sites (bottom panel of Fig. 2). This spatial separation of photoexcited hole and electron leads to a smaller oscillator strength, c.f. peak-I, suppresses the recombination rate and results in a longer lifetime, similar to interlayer exciton lifetimes in type-II heterojunctions[5,30]. Further, the distinct in-plane CT nature of peak-III makes it unlikely to relax to peak-I.

Dynamics of peak-II are anomalous in both devices; it undergoes a rapid initial decay (< 100 ps) followed by oscillations with periods of ~ 190 ps in D1 and ~ 890 ps in D2. We attribute this distinct behaviour of peak-II to its relatively larger extent in the moiré superlattice, possessing a hybrid character of partially in-plane CT and partially Wannier-like nature. For peak-II, some stacking sites have strong overlap between the photoexcited hole and electron: the photoexcited electron density is highest at AA site when hole is also at AA site (middle panel of Fig. 2). Conversely, when hole is at $B^{Se/W}$ site, electron density is spatially separated. In addition, exciton corresponding to peak-II has contributions from a larger number of single-particle valence and conduction states of the superlattice (c.f. peak-I), providing larger access to phonon scattering channels (Supplementary Section 9). This reduces its overall lifetime relative to peak-I and peak-III.

**Moiré intralayer intervalley biexciton**

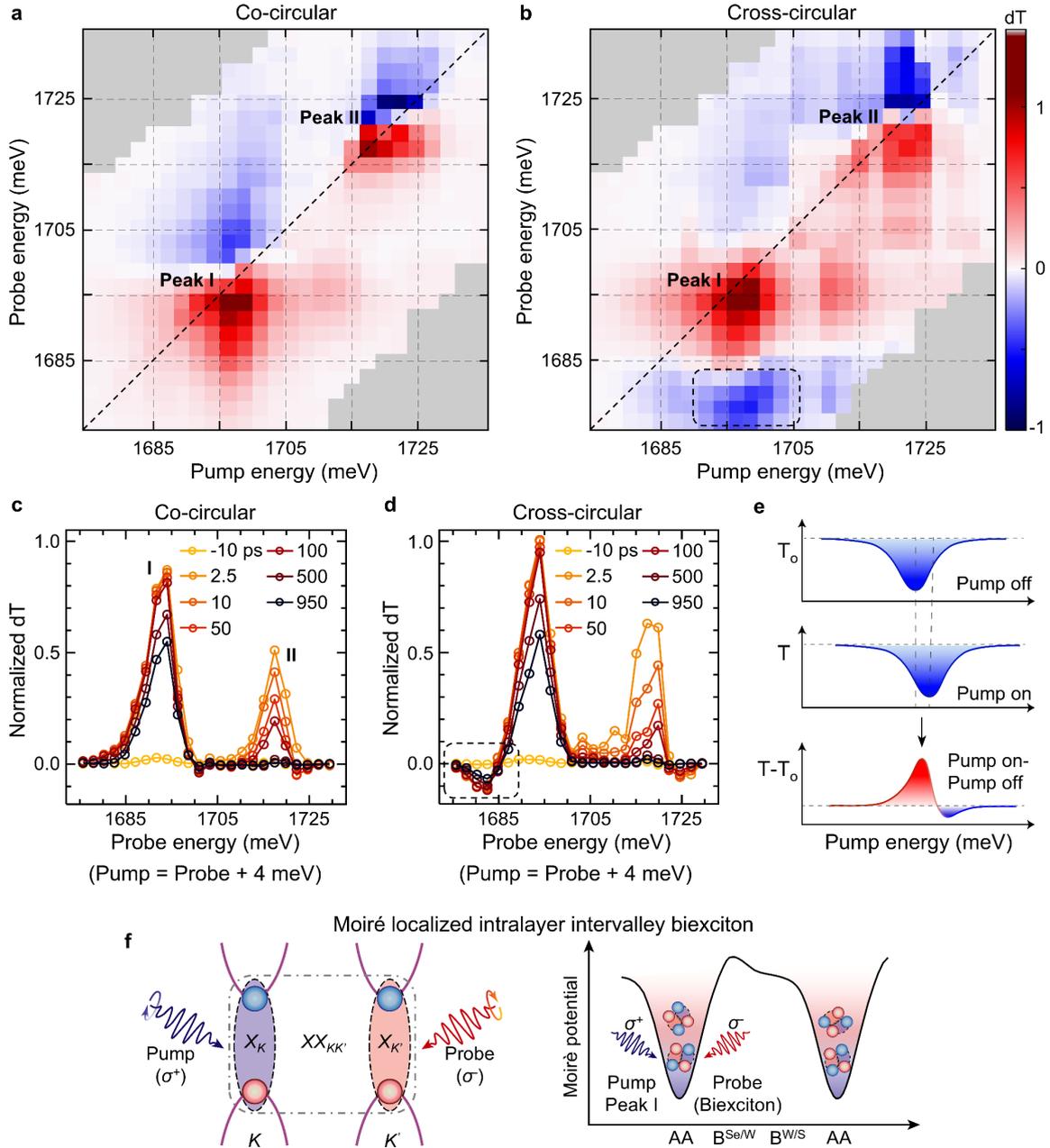

**Fig. 3: Observation of moiré intralayer intervalley biexciton. a, b,** The TCPP map of device D1 at 2.5 ps, for **(a)** co-, and **(b)** cross-circular polarization (CP). The intervalley biexciton feature is observed only in cross-CP

when pumped at resonance I and probed at energies lower than peak-I, as shown by the dotted box in (b). The colour scale has been saturated to enhance the visibility of weaker off-diagonal features. **c, d,** The line cuts along quasi-degenerate energies at various delay times for co- and cross-CP, respectively. In cross-CP, the biexciton exhibits a long lifetime and correlation with peak-I (in dotted box). **e,** Schematic of pump-induced spectral shifts in exciton resonance and resulting dT signals. **f,** Schematic depicts intralayer intervalley biexciton formation mechanism upon excitation using cross-CP pump and probe. $X_K$ and $X_{K'}$ denote the exciton in $K$ and $K'$ valleys, respectively. $XX_{KK'}$ is intralayer intervalley biexciton formed from the binding of excitons from both valleys, which are localized in the moiré potential.

To disentangle valley-specific phenomena from the population dynamics, we perform helicity-resolved TCPP measurements on device D1. Fig. 3a and 3b show co- ($\sigma^+/\sigma^+$) and cross- ($\sigma^+/\sigma^-$) circularly polarized (CP) transient dT maps, respectively, recorded at 2.5 ps delay. Relaxation dynamics of all features are observable from the line cuts along quasi-degenerate energies (where pump is 4 meV higher in energy than probe) at multiple time delays for co- and cross-CP in Fig. 3c and d, respectively. In cross-CP map, a distinct negative peak emerges when pumped at peak-I and probed at lower energies (dotted box in Fig. 3b). This peak is absent in steady-state dT measurements (Fig. 1c), indicating its many-body origin. Appearance of this signal exclusively under cross-circular excitation and probing, is a signature of the intervalley origin (Supplementary Section 10). This new peak has a long relaxation time (shown in dotted box, Fig. 3d and Supplementary Section 11), mirroring the primary Wannier exciton (peak-I), indicating they are correlated. Furthermore, this peak is extremely weak when pumped resonantly compared to when pumped at peak-I (Supplementary Section 10), highlighting peak-I as the primary source for this state.

This new state originating from many-body interactions and having a binding energy of ~ 16 meV in D1 and D2 (Supplementary Section 11), significantly lower than trion[34] binding energy, is identified as the moiré intralayer intervalley biexciton. These moiré biexcitons are formed from interaction of two IALX residing in opposite valleys ($K$ and $K'$) as illustrated schematically in Fig. 3f. While biexcitons have been reported in monolayer TMDs, they typically exhibit very fast relaxation[2–4]. Observing a stable, long-lived intervalley biexciton directly results from the moiré potential-induced localization of excitons (Fig. 3f), which suppresses their recombination pathways and fundamentally alters their many-body interactions. To our knowledge, this is the first observation of intralayer intervalley biexcitons in a moiré heterostructure. Their absence in previous studies may be attributed to the use of non-resonant excitation methods and ultrafast relaxation to IELX.

In addition to the primary bleaching signals, we also observe negative dT features at higher energies than peak-I and peak-II, which are attributed to transient pump-induced energy shifts in the exciton resonance, as illustrated in Fig. 3e. The observed asymmetry of these peaks in dT-maps, which would otherwise be symmetric, supports this interpretation. A more detailed analysis of these energy shifts is provided in Supplementary Section 12.

**Time domain signature of exciton-phason coupling in moiré IALX dynamics**

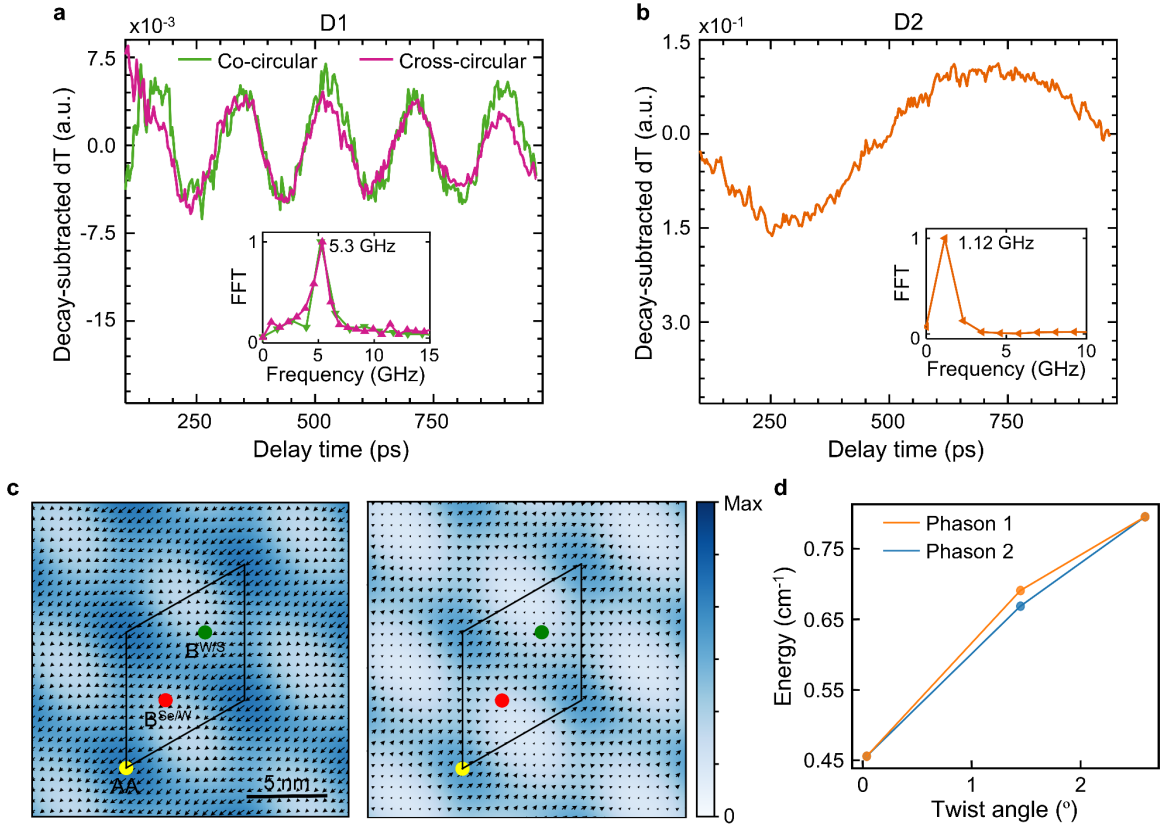

**Fig. 4: Observation of twist angle dependent exciton-phason coupling. a,** Comparison of oscillations extracted from co- and cross-CP pump-probe signals (after subtracting the exponential decay function) at the resonance of peak-II of device D1. The oscillations show similar amplitude and period (~ 190 ps) for both CP, indicating their helicity-independent nature. The Fourier transform in the inset shows that the frequency of these oscillations is ~ 5.3 GHz for both co- and cross-CP. **b,** The oscillations extracted from the LP pump-probe signal of device D2, at the resonance of peak-II, have a period of ~ 890 ps, and frequency ~ 1.12 GHz (shown in inset). **c,** Phason eigenmode component of $WS_2$ ($WSe_2$) on left (right), where arrows indicate the in-plane eigenvector direction, and their length and colour both represent its magnitude. For clarity, arrows are shown only for W atoms in the $WSe_2/WS_2$ layers, which are encapsulated by hBN. **d,** Shows the twist angle dependence of phason energy for a $WSe_2/WS_2$ heterostructure.

We now focus on the dT oscillations in pump-probe measurements of peak-II, which persist over the entire measured time-delay range (Fig. 1e and f). The oscillations in D1 (extracted after subtracting the exponential decay component, Supplementary Section 13) have a period of ~ 190 ps, and its Fourier transform shows a distinct peak at ~ 5.3 GHz (Fig. 4a, inset). Helicity-resolved TCPP show that these oscillations are invariant with excitation helicity (Fig. 4a), demonstrating that the collective mode responsible for these oscillations is not coupled to spin and valley degrees-of-freedom. Further, LP pump-probe signal of D2 also shows oscillations with an extended period of ~ 890 ps (~ 1.12 GHz, Fig. 4b).

These oscillations have twist angle-dependent frequency, very low energy (~ μeV), and no CP dependence. Our findings rule out moiré magnons, which are polarization sensitive[35], and plasmons, which have higher energy[36] (THz oscillations) (Supplementary Section 14). Instead, our evidence points to coupling of excitons with phasons, that are ultrasoft collective modes of moiré superlattice. To understand the origin of these phasons, we performed force-field based phonon calculations for hBN encapsulated $WSe_2/WS_2$ heterostructure (Supplementary Section 15 for details). Our calculations show two ultrasoft phason modes with energies below 1 cm$^{-1}$, which are strongly confined to $WSe_2/WS_2$ bilayer region. These modes are shear-like, involving opposite in-plane displacements of W atoms in the two TMD layers, with phonon eigenvectors sharply localized near AA-stacking site (Fig. 4c, Supplementary Video 1 and 2). This spatial localization is a direct manifestation of structural reconstructions in the moiré superlattice.

As moiré potential and exciton localization arise from structural reconstructions, particularly in-plane strain redistributions in $WSe_2$, these $WSe_2/WS_2$ phason modes constitute a direct dynamical channel for modulating the moiré potential. In other words, phasons dynamically modulate the effective moiré potential experienced by IALX that makes phasons observable through IALX dynamics. Exciton–phason coupling is expected only in the case of sufficient spatial overlap between phason modes and excitonic wavefunction[20]. Among the IALX, peak-II exhibits the most pronounced oscillations due to its spatially-extended excitonic wavefunction with significant overlap with the phason at AA stacking. Although small amplitude oscillations are present for most excitonic states, their amplitudes vary significantly (Supplementary Section 16). Surprisingly, peak-I shows smaller amplitude oscillations in-spite of its wavefunction overlap with the phason, which could result from reduced scattering channels due to lesser number of single-particle states that contribute to its exciton wavefunction (Supplementary Section 9). The extended nature of peak-II exciton allows it to effectively integrate over potential modulations across a larger area, resulting in enhanced coupling and stronger oscillations.

The lower phason energy in low-twist-angle D2, c.f. D1, is another piece of the puzzle. Theoretical studies have suggested that phason frequencies strongly depend on twist angle[17,19]. As twist angle is reduced, size of moiré unit cell increases, and the moiré potential landscape becomes smoother and flatter, reducing the potential energy barrier for interlayer sliding, and leading to softening of phason modes. To substantiate this, we performed phonon calculations and found a systematic increase in energies of these ultrasoft modes with increasing twist angle (Fig. 4d), which further validates our experimental results. The extremely low energy of phason modes poses a significant experimental challenge for conventional phonon spectroscopy, such as Raman spectroscopy. Pump-probe spectroscopy overcomes this fundamental limitation by operating in the time-domain.

**Conclusion and Outlook:**

In conclusion, we report the dynamics of moiré IALX by suppressing interlayer CT via excitation-rate engineering. We observe long lifetime of peak-I ~ 1000 ps (localized Wannier nature), and even longer lifetime of peak-III exciton (in-plane CT-character). Moiré intralayer intervalley biexcitons are observed with binding energy of ~ 16 meV, with exceptionally long lifetime due to moiré confinement, suggesting a potential source for entangled photons and quantum simulators[37]. Importantly, we present the first dynamical signatures of exciton-phason coupling via GHz oscillations in IALX dynamics. Twist-angle-dependent phason frequencies directly manifest the geometric control of dynamics via modifying the moiré potential. Observing phason and exciton modes in moiré superlattices is crucial because local lattice modulations can directly modulate electronic band structure[38,39].

This foundational insight into correlated excitonic and phononic dynamics is essential for future design of hybrid quantum devices, with applications envisioned in GHz optoelectronics and quantum sensing. Excitation rate engineering can be extended beyond 2D systems, such as to measure perovskites and quantum dots beyond low-energy defect and surface trap states. Importantly, our time-domain approach offers a method for exploring the dynamics of diverse low-energy collective modes in quantum materials. Thus, this work establishes moiré materials as unparalleled platforms for tuneable light-matter interactions.

**Methods:**

**Sample preparation:** Bulk crystals of $WS_2$ and $WSe_2$ were obtained from 2D Semiconductors. The hBN used for encapsulation was provided by the National Institute for Materials Science (NIMS), Japan. Monolayers of $WS_2$ and $WSe_2$ were mechanically exfoliated from their bulk crystals via Nitto blue tape, and few-layer hBN flakes were exfoliated using standard adhesive tape on $SiO_2$/Si substrate. The monolayer nature of the TMDs flakes was verified using a RAW imaging-based layer-identification technique[40]. Twisted moiré heterostructures were prepared using a standard dry-transfer method employing a polydimethylsiloxane/polycarbonate (PDMS/PC) stamp. The fabrication protocol involved the sequential pickup of individual layers from the $SiO_2$/Si substrates at a temperature of 110 °C, and twist between layers was controlled using a high precision rotational stage, followed by the final drop of the assembled stack on a double-sided polished sapphire substrate at 180 °C. The residual polymeric contaminants on the stacked devices were removed by a cleaning procedure involving sequential five-minute immersions in chloroform and isopropyl alcohol (IPA), followed by drying with nitrogen gas. Finally, the completed devices were subjected to thermal annealing at 250 °C for three hours within an inert-atmosphere glovebox to enhance the quality of the interlayer interface.

**Optical measurements:** All optical measurements were performed with the sample inside a closed-cycle Montana S-50 cryostat maintained at a temperature of 4 K. A custom-built, free-space optical setup was utilized for all experiments. For PL, a Ti-Sapphire pulsed laser (Mai Tai HP) was used, providing pulses of 100 fs duration at 80 MHz repetition-rate. The excitation beam was focused onto the sample through the cryostat window using a 50x long-working-distance objective (Mitutoyo, 0.42 NA). The PL signal was collected in transmission geometry and was subsequently directed to an Andor Kymera 328i spectrometer and detected with iDus 416 silicon CCD.

TRPL: In TRPL measurements, the spectrometer was operated as a monochromator to isolate a narrow spectral window of the emission. This spectrally filtered signal was then focused onto a single-photon avalanche diode (SPAD, Micro Photon Devices). Photon arrival times were recorded using a time-correlated single-photon counting (TCSPC) module (PicoQuant, PicoHarp 300). To probe decay dynamics over longer timescales, the laser repetition-rate was reduced from 80 MHz down to 800 kHz by employing an external pulse picker (Spectra-Physics, Model 3980).

SHG: SHG measurements were conducted using 820 nm pulsed laser output at 80 MHz for excitation, with the SHG signal at 410 nm collected in transmission geometry.

White light transmission: A halogen lamp (Amscope) served as a broadband white-light source for steady-state transmission. The lamp output was fiber-coupled and spatially filtered through a 100 μm pinhole and focused onto the sample through an objective, resulting in a beam diameter of ~ 2.5 μm. The transmitted light traversed the same optical path as the PL collection. The transmittance contrast was calculated using the formula, $dT/T_o = (T-T_o)/T_o$, where $T$ and $T_o$ represent the transmitted intensity through the heterostructure and the bare substrate, respectively.

Pump probe spectroscopy: The ultrafast optical response was investigated using a custom-built TCPP spectroscopy in a transmission geometry. The output of the Ti-Sapphire laser was split into pump and probe beams that are spectrally shaped independently, using grating-based pulse-shapers, generating ~ 1 nm full width at half maxima (FWHM) pulses. A mechanical stage in the pump path provided a precise temporal delay relative to the probe pulse. Both the beams were modulated at different frequencies ($\Omega_1$ and $Ω_2$) using acoustic optical modulators. Independent polarization control for each beam was achieved using a combination of a half-wave plate and a linear polarizer. The two beams were then recombined, collinearly propagated, and focused onto the sample using a 50x objective with spots of ~ 2.5 μm (pump) and 2.0 μm (probe), ensuring uniform photoexcitation in the probed region. For helicity-resolved experiments, a quarter-wave plate was inserted into the common path before the objective. The transmitted probe signal was isolated from the pump via cross-polarization or spectral filtering and measured with a photodetector. Finally, the signal was demodulated (at reference frequency, $Ω_1 - Ω_2$) using a lock-in amplifier (Stanford Research Systems, SRS-830) to extract the transient differential transmission signal.

**Theoretical Methods:**

Phonon calculations: We use a rectangular moiré unit cell of size 142 × 164 Å. Additionally, we place a 32 × 32 orthorhombic hBN supercell on both sides of the $WSe_2$/$WS_2$ bilayer. Structural relaxation of the moiré superlattice is performed using classical force-field calculations using Large-scale Atomic/Molecular Massively Parallel Simulator (LAMMPS)[41] . Intralayer interactions within TMD layers are described by the Stillinger–Weber potential[42], and for hBN with Tersoff[43], while interlayer coupling between adjacent layers is modelled using the Kolmogorov–Crespi potential[44]. Atomic positions are relaxed until the maximum force on any atom falls below $10^{-6}$ eV/Å. The force-constant generation and phonon calculations for the moiré system are carried out using the PARPHOM package (Supplementary Section15 for more details).

Moiré exciton calculations: We performed density functional theory (DFT) calculations to investigate the electronic structure of the reconstructed $WSe_2$ layer in a 0°-aligned $WS_2$/$WSe_2$ heterostructure. The heterostructure was structurally relaxed using the classical force fields described above. Subsequent GW-BSE calculations were carried out using the pristine unit-cell matrix projection (PUMP) method to compute the IALX spectrum and to elucidate the nature of the moiré excitons responsible for the excitonic peaks in the absorption spectrum. The twist-angle dependence of the exciton wavefunctions and optical absorption was further examined using a continuum model. Full details of the structural relaxations, DFT and GW-BSE calculations, as well as the continuum model, are provided in the Supplementary Section 8 and 9.

**Other sections:** Supplementary information is included with this manuscript.

**Author information:**

Corresponding Author(s): *Mit H Naik, mit.naik@austin.utexas.edu, *Manish Jain, mjain@iisc.ac.in, *Akshay Singh, aksy@iisc.ac.in

**Author Contributions:**

RD and AS developed the experimental framework, with assistance from H. Swaminathan. R. Dutta and MHN performed the GW-BSE calculations using the PUMP method to study the nature of the moiré excitons in the 0° aligned $WS_2/WSe_2$ heterostructure. HS and MJ computed the twist-angle dependence of the moiré exciton spectrum using a continuum model and carried out phonon calculations to study the phasons and their localization in the moiré superlattice. RD performed the experiments, and data analysis with guidance from AS. RD and AS discussed and prepared the manuscript, with contributions from all authors. All authors have given approval to the final version of the manuscript.

**Data Availability**

All data is available upon reasonable request.

**Acknowledgments**

AS and MJ acknowledge funding from Department of Science and Technology (DST) Nanomission CONCEPT grant (NM/TUE/QM-10/2019). AS acknowledges funding from the Ministry of Education through the MoE-STARS (STARS-2/2023-0265), and Anusandhan National Research Foundation grant (CRG-2022-003334). MHN acknowledges support by the National Science Foundation (NSF) through the Center for Dynamics and Control of Materials: an NSF Materials Research Science and Engineering Center (Cooperative Agreement No. DMR-2308817). R. Dutta was supported by the Texas Quantum Institute Graduate Fellowship. H.S. acknowledges financial support from the University Grants Commission through UGC-NET program.

# Signatures of moiré intralayer biexcitons and exciton-phason coupling in WSe$_2$/WS$_2$ heterostructures

Ranju Dalal[1], Harsimran Singh[1], Rwik Dutta[2], Hariharan Swaminathan[1], Kenji Watanabe[3], Takashi Taniguchi[4], Mit H Naik[2*], Manish Jain[1*], Akshay Singh[1*]

[1]*Department of Physics, Indian Institute of Science, Bengaluru, Karnataka 560012, India*

[2]*Department of Physics and Center for Complex Quantum Systems, University of Texas at Austin, Austin, Texas 78712, USA*

[3]*Research Center for Functional Materials, National Institute for Materials Science, Ibaraki 3050044, Japan*

[4]*International Center for Materials Nanoarchitectonics, National Institute for Materials Science, Ibaraki 3050044, Japan*

*Corresponding author: mit.naik@austin.utexas.edu, mjain@iisc.ac.in, aksy@iisc.ac.in

**Sections**:

1. Identification of stacking type using Second Harmonic Generation (SHG)
2. Optical image of samples
3. Photoluminescence (PL) of device D2
4. Comparison of transmission contrasts of D1, D2, and WSe$_2$ monolayer
5. Two-colour pump-probe (TCPP) setup
6. Excitation repetition rate-dependent time-resolved PL (TRPL) on Device D2
7. Literature reports on TCPP measurements at lower excitation repetition-rate
8. Moiré excitons in 0° WS$_2$/WSe$_2$ superlattice
9. Twist angle dependence of exciton wavefunction and optical absorption spectrum
10. Correlation between peak I and biexciton
11. Dynamics of biexcitonic state
12. Peak shifts in TCPP maps
13. Raw data and decay subtraction of peak II to extract oscillations
14. Identification of GHz oscillations as phasons
15. Phason modes for hBN encapsulated WSe$_2$/WS$_2$
16. GHz oscillations in other excitonic states

## Section 1: Identification of stacking type using Second Harmonic Generation (SHG)

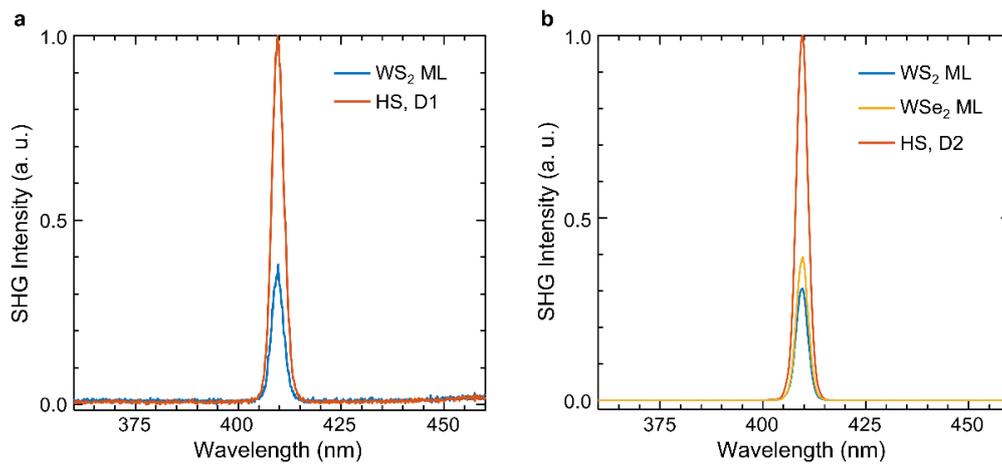

**Fig. S1:** **(a)** and **(b)** show the stronger SHG signal from heterostructure compared to $WSe_2$ and $WS_2$ monolayers (ML), confirming R-type stacking of both devices D1 and D2, respectively, matching with the previous reports[1].

## Section 2: Optical image of samples

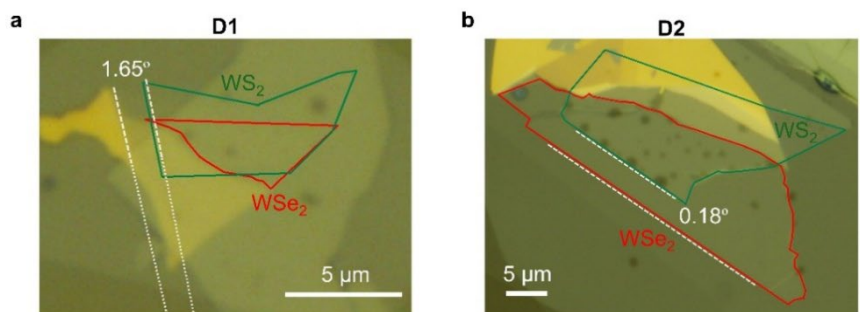

**Fig. S2:** **(a)** and **(b)** show the optical microscopic image of device D1 and D2, highlighting $WSe_2$, $WS_2$ MLs and the approximate twist angle of 1.65º and 0.18º, respectively.

## Section 3: Photoluminescence (PL) of device D2

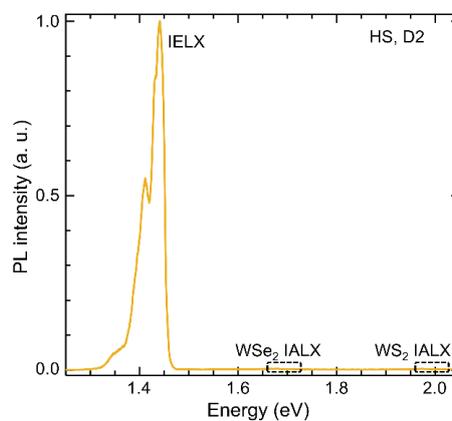

**Fig. S3:** PL of device D2, highlights strong interlayer excitons (IELX) emission and highly quenched intralayer excitons (IALX), which signifies the strong electronic coupling between the layers.

## Section 4: Comparison of transmission contrasts of D1, D2, and WSe$_2$ monolayer

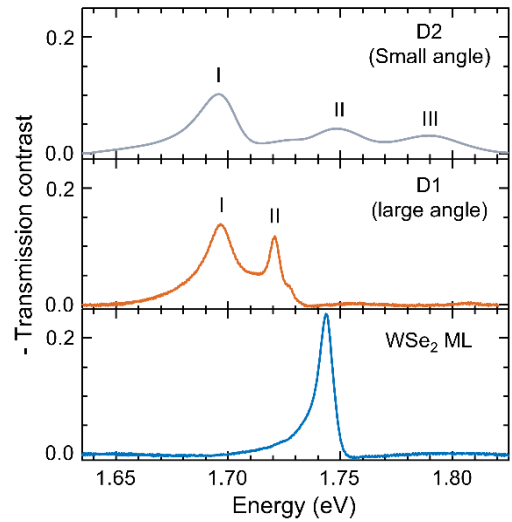

**Fig. S4:** WSe$_2$ monolayer (ML) shows a single resonance around 1.74 eV, whereas D1 and D2 show two and three states, respectively, near the A-exciton resonance of WSe$_2$ in the moiré heterostructures, highlighting the effective modulation of IALX states by the strong, periodic moiré potential.

## Section 5: Two-colour pump-probe (TCPP) setup

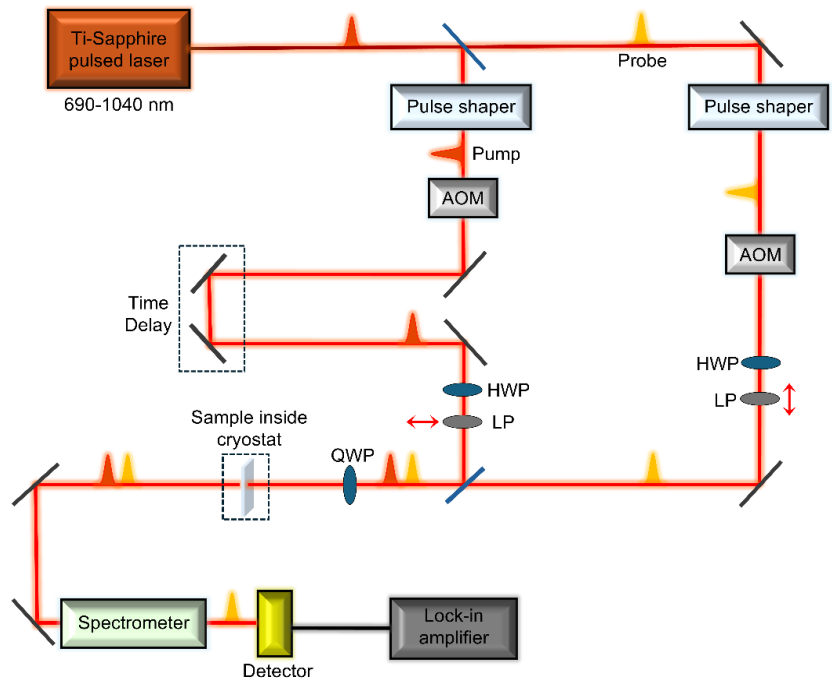

**Fig. S5:** Schematic of TCPP experimental setup used in our measurements.

**Section 6: Excitation repetition rate-dependent time-resolved PL (TRPL) on Device D2**

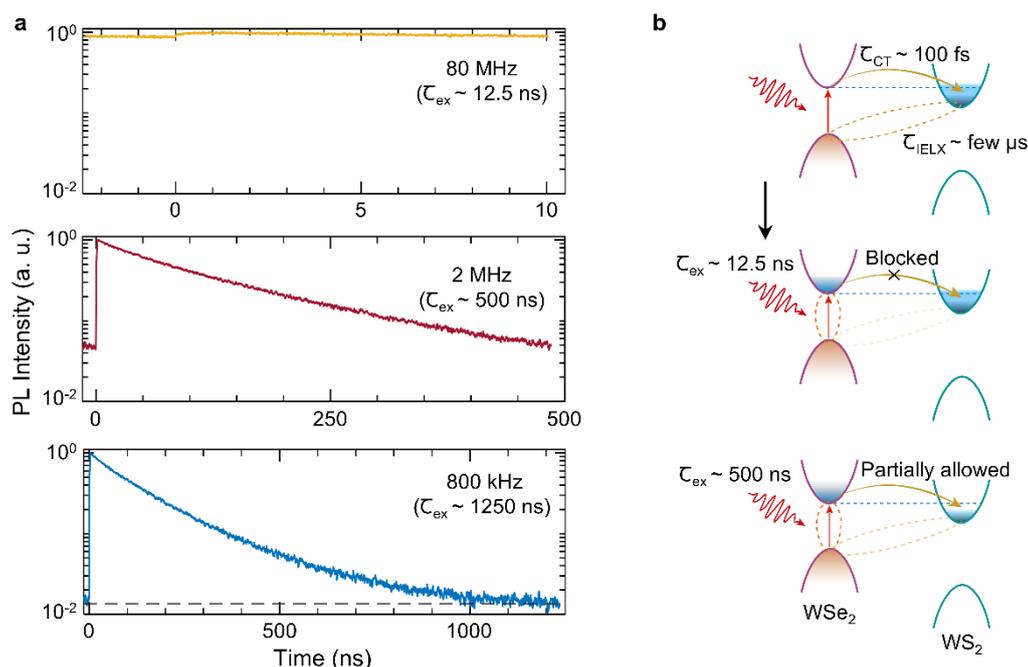

**Fig. S6: (a)** Time-resolved PL (TRPL) of D2 shows very long-lived IELX, which are saturated by 80 MHz excitation rate. Their saturation decreases with decreasing excitation rate as IELX partially decays before the next laser pulse is incident on the sample. The dotted line (in bottom panel) represents the dark counts of the TRPL system. The comparative dark counts for 80 MHz and 2 MHz are very low. **(b)** depicts the blocked (middle) and partially allowed (bottom) relaxation pathways at higher and lower repetition rates, respectively.

TRPL on device D2 (Fig. S6a) shows long-lived IELX (> 500 ns), creating a persistent population background under high excitation rate (80 MHz) because they cannot fully decay between consecutive pulses. This background reaches the dark counts level at a lower excitation rate (800 kHz), indicating complete decay of IELX.

This understanding provides a powerful experimental strategy for measuring the IALX relaxation directly to the ground state. By operating at a high repetition rate (80 MHz), we intentionally create a saturated, steady-state IELX population. This blocks the CT channel from IALX to IELX which would otherwise be partially open at lower repetition rate (Fig. S6b). This approach effectively decouples the IALX dynamics from the IELX conversion pathway, which allows the measurement of IALX properties, resolving the prior experimental ambiguities in their lifetime due to variable CT efficiency.

**Section 7: Literature reports on TCPP measurements at lower excitation repetition-rate**

The ultrafast charge transfer (CT) from one layer to another depends on various factors, most commonly, interlayer coupling and excitation repetition rate. Repetition rate can decide whether the CT channel would be blocked (at high rate) or partially available (at lower rate) and hence controls the effect of IELX on IALX dynamics. In the literature, various reports show long-lived dynamics at the resonance of IALX in TCPP attributed to IELX, considering that IALX cannot be so long-lived. However, all those measurements are at lower repetition rates, which gives IALX and IELX enough time to relax; hence, the results show their mixed dynamics. Thus, measurements at low repetition rates are insufficient to probe the dynamics of IALX.

The following table shows some of these reports:

| Heterostructure | Pump and Probe energies | Repetition rate | References |
|---|---|---|---|
| $WSe_2/WS_2$ | Pump at 1.75 eV and white-light probe | 250 kHz | 2 |
| $WSe_2/WS_2$ | Pump at 3.10 eV and probe $WSe_2$ and $WS_2$ resonances | 400 kHz | 3 |
| $WSe_2/WS_2$ | Pump at 90 meV above $WSe_2$ A exciton resonance and white light probe | Yb:KGW laser (kHz) | 4 |
| $WSe_2/MoS_2$ | Pump at $WSe_2$ resonance and probe $WSe_2$ and $MoS_2$ resonances | 250 kHz | 5 |

**Section 8: Moiré excitons in 0° $WS_2/WSe_2$ superlattice**

**Structural reconstructions:** The initial structure of the rotationally aligned $WS_2/WSe_2$ superlattice is constructed with the help of TWISTER code[6]. A $25 \times 25$ in-plane superlattice of $WSe_2$ is commensurate with a $26 \times 26$ in-plane superlattice of $WS_2$ at a zero twist-angle between the layers, creating a moiré pattern with 8 nm periodicity. The moiré superlattice structure is relaxed using reliable classical force fields parameterized against van der Waals-corrected density functional theory (DFT) data. Intralayer bonding is modelled with the Stillinger–Weber potential[7], while interlayer interactions are captured using the Kolmogorov–Crespi potential[8,9], both implemented within the LAMMPS software package[10].

**Density functional theory calculations:** Fig. S7 shows the DFT band structure of reconstructed $WSe_2$ superlattice and the modulation of electron charge density of the bands close to the valence and conduction band edges at the $\gamma$ point in moiré Brillouin zone.

The single-particle electronic structure of the reconstructed $WSe_2$ superlattice is computed using DFT[11]. The calculations are performed with the SIESTA code[12], employing the Perdew–Burke–Ernzerhof (PBE) generalized gradient approximation[13] for exchange-correlation effects and norm-conserving pseudopotentials[14,15]. Spin-orbit coupling is not included in the calculations. To obtain the self-consistent ground-state charge density, the moiré Brillouin zone is sampled only at the zone-center $\gamma$ point. After self-consistency is achieved, the DFT Hamiltonian is evaluated at selected k-points in the moiré Brillouin zone to compute the electronic band structure and wavefunctions used in subsequent excited-state calculations.

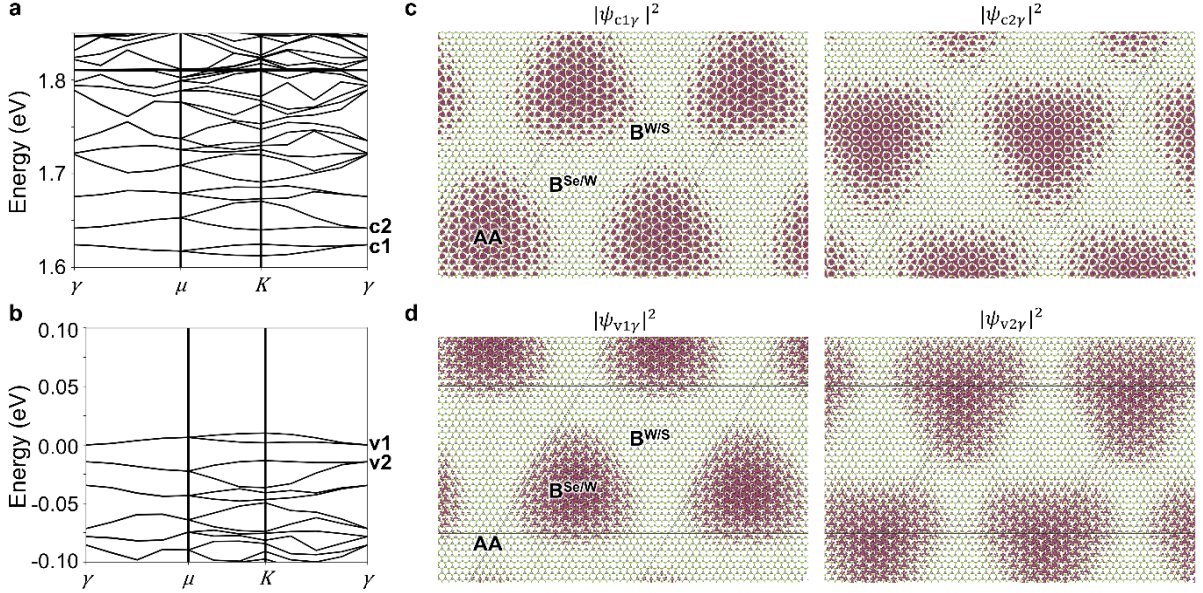

**Fig. S7: (a)** and **(b)** DFT band structures at the conduction and valence band edges of the reconstructed WSe$_2$ superlattice, respectively. **(c)** and **(d)** conduction and valence charge densities plotted at the $\gamma$ point in the moiré Brillouin zone, for the corresponding band states marked in (a) and (b), respectively.

Calculations for the pristine unit cells are carried out using the plane-wave-based Quantum ESPRESSO package[16]. The electronic wavefunctions are expanded in a plane-wave basis with a kinetic-energy cutoff of 40 Rydberg (Ry). For consistency across computational approaches, identical exchange–correlation functionals and pseudopotentials are used in both the SIESTA and Quantum ESPRESSO calculations.

**Pristine unit-cell matrix projection (PUMP) method:** The Bethe-Salpeter equation (BSE) for the moiré superlattice, to compute the moiré excitons and absorption spectrum, is solved using the pristine unit-cell matrix projection (PUMP) method[17,18]. The excitonic eigenvalue problem is written as

$$\left(E_{c\mathbf{k}}^{QP} - E_{v\mathbf{k}}^{QP}\right)A_{cv\mathbf{k}}^{S} + \sum_{v'c'\mathbf{k}'} \langle \psi_{v\mathbf{k}}^{SL}\psi_{c\mathbf{k}}^{SL}|K^{eh}|\psi_{v'\mathbf{k}'}^{SL}\psi_{c'\mathbf{k}'}^{SL}\rangle A_{c'v'\mathbf{k}'}^{S} = \Omega_S A_{cv\mathbf{k}}^{S}, \quad (1)$$

where $E_{n\mathbf{k}}^{QP}$ denote quasiparticle energies, $\psi_{n\mathbf{k}}^{SL}$ are superlattice wavefunctions, and $A_{cv\mathbf{k}}^{S}$ and $\Omega_S$ are the exciton eigenvectors and eigenvalues, respectively. The electron-hole interaction kernel $K^{eh}$ consists of an attractive screened direct term and a repulsive exchange term. Direct evaluation of the kernel matrix elements represents the dominant computational cost of BSE calculations for large moiré superlattices.

The PUMP method reduces this cost through a two-step procedure. First, the electronic wavefunctions of the WSe$_2$ superlattice with the strain redistributions are expanded in a basis of pristine primitive unit-cell states. Each valence and conduction superlattice wavefunction is expressed as a linear combination of pristine unit-cell valence and conduction wavefunctions, respectively: $|\psi_{v\mathbf{k}}^{SL}\rangle = \sum_i a_i^{v\mathbf{k}}|\Phi_{i\mathbf{k}}^{val}\rangle$,

$|\psi_{c\mathbf{k}}^{SL}\rangle = \sum_i a_i^{c\mathbf{k}} |\Phi_{i\mathbf{k}}^{cond}\rangle$, where $|\Phi_{i\mathbf{k}}^{val}\rangle$ and $|\Phi_{i\mathbf{k}}^{cond}\rangle$ denote pristine valence and conduction states of the pristine superlattice, respectively. These pristine superlattice states are directly related to states in the primitive unit-cell Brillouin zone through band folding.

In the second step, the electron–hole interaction kernel matrix elements of the WSe$_2$ superlattice are approximated as coherent linear combinations of pristine unit-cell kernel matrix elements using the expansion coefficients $a_i^{v\mathbf{k}}$ and $a_i^{c\mathbf{k}}$. Specifically, each superlattice kernel element is written as

$$\langle \psi_{v\mathbf{k}}^{SL} \psi_{c\mathbf{k}}^{SL} | K^{eh} | \psi_{v'\mathbf{k}'}^{SL} \psi_{c'\mathbf{k}'}^{SL} \rangle$$

$$\approx \sum_{ijpq} a_i^{v\mathbf{k}*} a_j^{c\mathbf{k}*} a_p^{v\mathbf{k}} a_q^{c\mathbf{k}} \langle \Phi_{i\mathbf{k}_m}^{val} \Phi_{j\mathbf{k}_m}^{cond} | K^{eh} | \Phi_{p\mathbf{k}_m'}^{val} \Phi_{q\mathbf{k}_m'}^{cond} \rangle, \quad (2)$$

$$\approx \sum_{ijpq} a_i^{v\mathbf{k}*} a_j^{c\mathbf{k}*} a_p^{v\mathbf{k}} a_q^{c\mathbf{k}} \langle \phi_{s\mathbf{k}_{uc}^1}^{val} \phi_{t\mathbf{k}_{uc}^2}^{cond} | K^{eh} | \phi_{y\mathbf{k}_{uc}^3}^{val} \phi_{z\mathbf{k}_{uc}^4}^{cond} \rangle \delta_{\mathbf{k}_{uc}^3-\mathbf{k}_{uc}^1, \mathbf{k}_{uc}^4-\mathbf{k}_{uc}^2}, \quad (3)$$

here, $i, p$ label pristine valence bands and $j, q$ label pristine conduction bands, while the momenta in the superlattice Brillouin zone ($\mathbf{k}_m$) are related to those in the primitive unit-cell Brillouin zone ($\mathbf{k}_{uc}$) via band folding. Equation (2) corresponds to kernel matrix elements evaluated in the pristine supercell representation, whereas Eq. (3) involves kernel elements computed directly in the primitive unit cell. Screening contributions from both WSe$_2$ and WS$_2$ layers are included in the evaluation of the direct electron-hole interaction. By expressing each superlattice kernel matrix element as a linear combination of pristine unit-cell kernel elements, the PUMP approach enables a dramatic reduction in computational cost while retaining quantitative accuracy for excitonic properties of large moiré systems.

**Details of the GW-BSE calculations:** At the DFT level, the band gap is underestimated, although the overall band dispersion is reasonably well captured. To obtain accurate quasiparticle energies, the electron self-energy is evaluated within the GW approximation[19] for a monolayer WSe$_2$ in the pristine unit cell. The resulting GW correction to the band gap at the K point is then applied as a rigid shift to the DFT band structure of the moiré superlattice. All GW calculations are performed using the BerkeleyGW package[20]. The static dielectric function and screened Coulomb interaction are computed within the random phase approximation using a plane-wave cutoff of 35 Ry and approximately 5000 unoccupied states[19]. Frequency dependence of the dielectric response is treated using the Hybertsen-Louie generalized plasmon-pole model. To remove spurious interactions between periodic images, the Coulomb interaction is truncated along the out-of-plane direction[21]. Convergence with respect to Brillouin-zone sampling is improved using the nonuniform neck subsampling technique[22].

Excitonic effects are incorporated by solving the BSE using the BerkeleyGW package. The electron-hole interaction kernel for the moiré superlattice is constructed using the PUMP approach (described in detail above), in which the moiré kernel is expressed as a coherent linear combination of pristine unit-cell kernel matrix elements. For the WSe$_2$ layer with inhomogeneous strain reconstruction, the BSE Hamiltonian is built using 12 valence and 12 conduction moiré bands, with a $6 \times 6 \times 1$ k-point sampling of the moiré Brillouin zone. The moiré valence and conduction wavefunctions are expanded in terms of 24 pristine valence and 24 pristine conduction band super cell wavefunctions, respectively[17].

The static dielectric matrix used to compute the pristine unit-cell kernel matrix elements is calculated for the $WS_2$/$WSe_2$ AA stacking in the primitive unit cell of $WSe_2$, with the interlayer separation set to 6.3 Å. The dielectric matrix is expanded in plane waves up to a cutoff of 4.5 Ry and includes ~ 500 unoccupied states. The resulting unit-cell kernel matrix elements are used to assemble the moiré BSE Hamiltonian, which is diagonalized to obtain the imaginary part of the dielectric function, $\varepsilon_2(\omega)$. In the weak-absorption limit, the optical absorption is given by $A(\omega) = \frac{\epsilon_2(\omega)\omega d}{c}$, where $d$ is the effective sample thickness, taken to be the bilayer thickness of 9.5 Å.

Fig. S8 plots the GW-BSE absorption spectrum of the reconstructed $WSe_2$ layer showing the emergence of three moiré excitons peaks due to strain redistributions in the layer. The nature of each of the excitonic peaks is shown in the main text.

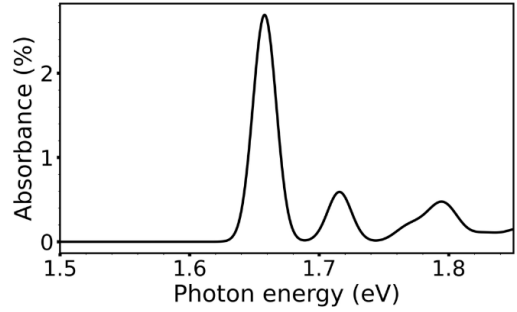

**Fig. S8:** GW-BSE absorbance spectrum of the reconstructed $WSe_2$ superlattice computed using the PUMP method.

**Section 9: Twist angle dependence of exciton wavefunction and optical absorption spectrum**

To elucidate the dependence of exciton energies and envelope function on twist angle, we numerically solve the Bethe–Salpeter equation (BSE) using a model single-particle Hamiltonian[23,24] that captures the essential band dispersions and Bloch eigenstates given as:

$$\widehat{H} = \hbar v_F \big(\tau k_x \widehat{\sigma_x} + k_y \widehat{\sigma_y}\big) \otimes \widehat{\mathcal{I}}_s + \frac{\Delta}{2}\widehat{\sigma_z} \otimes \widehat{\mathcal{I}}_s - \lambda\tau \frac{\widehat{\sigma_z}-1}{2} \otimes \widehat{s}_z. \quad (4)$$

Here, $\widehat{\sigma} = \big(\widehat{\sigma_x}, \widehat{\sigma_y}, \widehat{\sigma_z}\big)$ are Pauli matrices that acts on the orbital subspace spanned by $|d_{z^2}\rangle$ and $\frac{1}{\sqrt{2}}(|d_{x^2-y^2}\rangle + i\tau|d_{xy}\rangle)$, while $\widehat{\mathcal{I}}_s$ is the identity operator and $\widehat{s}_z$ is the Pauli spin operator acting on the spin subspace of spin ½ electron. The valley index $\tau = \pm 1$ distinguishes K and K' valleys, and $\Delta$ denotes the band gap at these high-symmetry points in the absence of spin–orbit coupling.

The first term, proportional to $\hbar v_F$, describes the Dirac-like kinetic coupling between orbital components with Fermi velocity $v_F$, yielding a linear band dispersion near the K/K' points. The second term, proportional to $\frac{\Delta}{2}\widehat{\sigma_z}$, introduces a mass gap between the conduction and valence bands, while the third term, proportional to $\lambda$, accounts for the intrinsic spin–orbit coupling that induces a valley-dependent spin splitting, where $\lambda$ controls the strength of spin-orbit coupling in the model.

We restrict our analysis to the bright A exciton, which arise from transitions between the top valence band and the bottom conduction band with identical spin projection $s_z = 1(-1)$ for K(K') valley. Accordingly, the effective Hamiltonian for this subspace, incorporating spin–orbit coupling, can be expressed as[24]

$$\widehat{H} = \hbar v_F\bigl(\tau k_x \widehat{\sigma_x} + k_y \widehat{\sigma_y}\bigr) + \frac{\Delta'}{2}\widehat{\sigma_z} \tag{5}$$

In addition, we assume that the effect of presence of $WS_2$ on $WSe_2$ can be modelled by an effective potential given as

$$V_m(\mathbf{r}) = \sum_\alpha \widetilde{V}_\alpha \, e^{i\mathbf{G}_m^\alpha \cdot \mathbf{r}} \tag{6}$$

Here, $\alpha$ labels the six first-shell reciprocal vectors, and $\mathbf{G}_m^\alpha$ denotes the moiré superlattice reciprocal vectors associated with a given $\alpha$. The subset $\alpha = 1, 3, 5$ is related by 120° rotational symmetry. Here $V_m$ is a 2x2 matrix for a given position. We compute the moiré unit cell lattice vectors by assuming the lattice constant ratio of 25/26 for $a_{WSe2}/a_{WS2}$ with twist angle $\theta$. In the present analysis, we restrict the summation to the six shortest (first shell) moiré reciprocal lattice vectors. The symmetry of the lattice imposes the following constraint:

$$\widetilde{V_1} = \widetilde{V_3} = \widetilde{V_5} = \widetilde{V_2^*} = \widetilde{V_4^*} = \widetilde{V_6^*} \tag{7}$$

where

$$\widetilde{V_1} = \begin{pmatrix} V_c e^{i\phi_c} & 0 \\ 0 & V_v e^{i\phi_v} \end{pmatrix} \tag{8}$$

Here $V_c$ and $V_v$ are the parameters that controls the strength of moiré potential, whereas $\phi_c$ and $\phi_v$ phase parameters control the localization of moiré potential. Single particle wavefunctions, after diagonalizing the Hamiltonian with moiré potential, is expressed as:

$$\psi_{n,\mathbf{k}}(\mathbf{r}) = \sum_{\mathbf{G}} U_{\mathbf{k},n}(\mathbf{G}) e^{i(\mathbf{k}+\mathbf{G})\cdot \mathbf{r}} \tag{9}$$

where $U_{\mathbf{k},n}$ is the eigenfunction obtained by diagonalizing the continuum Hamiltonian. At zero centre-of-mass momentum ($\mathbf{Q} = 0$), the excitonic Hamiltonian takes the form

$$\mathcal{H}^{exc}_{vc\mathbf{k},v'c'\mathbf{k}'} = (E_{c\mathbf{k}} - E_{v\mathbf{k}})\delta_{vv'}\delta_{cc'}\delta_{\mathbf{k}\mathbf{k}'} + \mathcal{V}^x_{vc\mathbf{k},v'c'\mathbf{k}'} - \mathcal{W}^d_{vc\mathbf{k},v'c'\mathbf{k}'} \tag{10}$$

where $v$ and $c$ denote valence and conduction bands, respectively. To simplify the expressions, we introduce the overlap matrix elements.

$$M^{(\mathbf{G})}_{i,j}(\mathbf{k},\mathbf{q}) = \sum_{\mathbf{G}'} U^*_{\mathbf{k}+\mathbf{q},i}(\mathbf{G}' + \mathbf{G}) U_{\mathbf{k},j}(\mathbf{G}') \tag{11}$$

which describe the overlap between the envelope functions of bands $i$ and $j$ under a reciprocal-lattice shift $\bm{G}$ and momentum transfer $\bm{q}$. Then, the direct (screened) interaction can be written as

$$\mathcal{W}^{d}_{vck,v'c'k'} = \frac{1}{A}\sum_{\bm{G}} W_{RK}(|\bm{q}+\bm{G}|) M^{(G)}_{c,c'}(\bm{k},\bm{q}) \left[M^{(G)}_{v,v'}(\bm{k}',\bm{q})\right]^*, \quad (12)$$

where A is the system area and $W_{RK}(\bm{q}) = \frac{e^2}{\epsilon_0}\frac{2\pi}{q(\kappa+r_0 q)}$ is the Rytova-Keldysh[25] screened Coulomb potential in momentum space. We take the value of $\kappa = 4.0$, $r_0 = 45$ Å (Ref.[26]). Similarly, the exchange term can be written as:

$$\mathcal{V}^{x}_{vck,v'c'k'} = \frac{\delta_{kk'}}{A}\sum_{\bm{G}\neq 0} V(|\bm{G}|) M^{(G)}_{c,v}(\bm{k},0) \left[M^{(G)}_{c',v'}(\bm{k},0)\right]^* \quad (13)$$

Where $V(|\bm{G}|) = \frac{e^2}{\epsilon_0}\frac{2\pi}{q}$ denotes the bare Coulomb interaction. We finally solve for the eigenvalue equation as

$$\sum_{v'c'k'} \mathcal{H}^{exc}_{vck,v'c'k'} A^{S}_{v'c'k'} = \Omega_S A^{S}_{vck} \quad (14)$$

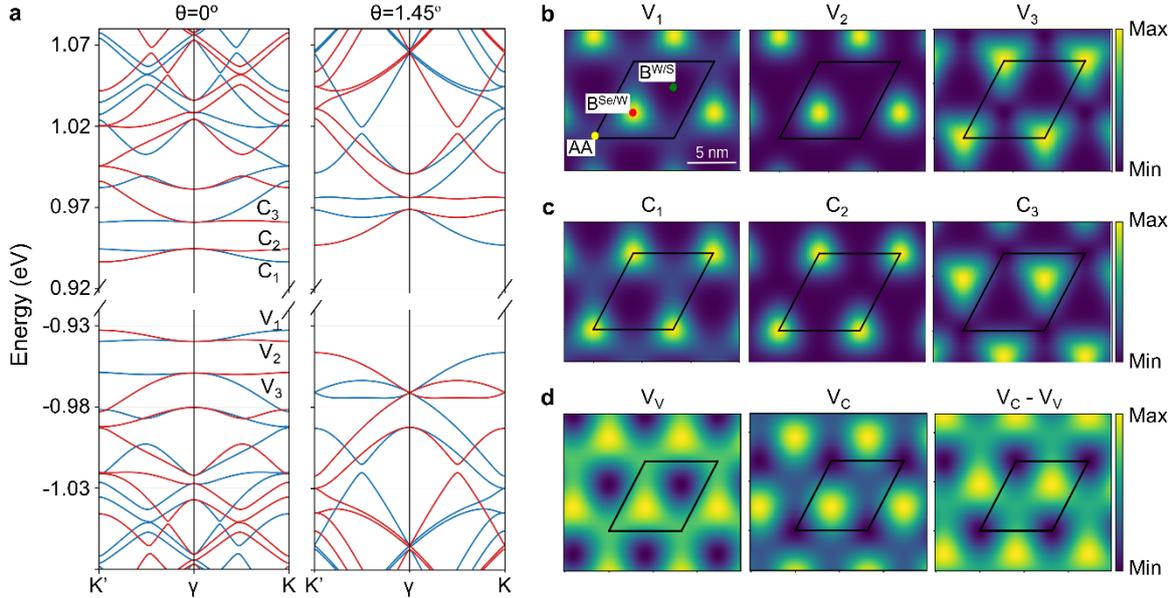

**Fig. S9:** (a) Band structures of moiré WSe$_2$/WS$_2$ bilayers at twist angles of 0° (left) and 1.45° (right), where red (blue) lines corresponds to K'(K) valley; (b) Density corresponding to single-particle valence-band wavefunctions at the K point for bands v1–v3 (left to right) for twist angle 0°; (c) Charge density corresponding to single-particle conduction-band wavefunctions at the K point for bands c1–c3 (left to right) for twist angle 0°; (d) real-space distributions of the moiré $V_V=V^{22}{}_m(\bm{r})$ and conduction-band potentials $V_C=V^{11}{}_m(\bm{r})$, (see eq. 6), along with the spatial profile of band gap ($V_C$-$V_V$) for twist angle 0°.

We fix $\phi_c = 45°$ and value of $\hbar v_F = 3.94$ eV consistent with Ref.[23] and choose $V_c = -14$ meV, $V_v = -16$ meV. The calculated band structure for twist angles 0° (left) and 1.45° (right) highlights the strong band reconstruction at 0°, as shown in Fig. S9. As reported in Ref.[27], lattice reconstruction at small twist angles enhances the localization of hole states near B$^{Se/W}$ site. Within our model, an analogous localization is obtained when the valence-band moiré potential is parameterized such that its maximum is centred at $B^{Se/W}$ site, which is controlled by the phase $\phi_v$.

We find that choosing $\phi_v = 105°$ results in strong localization of the hole envelope function at the $B^{Se/W}$ stacking site (see Fig. S9). With this phase, the calculated moiré band structure and the corresponding single-particle envelope functions exhibit close quantitative agreement with previously reported the *ab-initio* results[17], as well as those shown in Fig. S7. Both the band structure and the real-space localization characteristics of the single particle envelope function are accurately reproduced (see Fig. S9). For completeness, the spatial profiles of the moiré potential corresponding to this parameter choice are presented in Fig. S9.

Furthermore, consistent with Ref.[28], the influence of lattice reconstruction is expected to diminish with increasing twist angle. To capture this trend, we reduce the moiré-potential strength linearly to zero as the twist angle increases from 0° to 3°, while simultaneously sweeping $\phi_v$ from 105° to 135°. The choice $\phi_v = 135°$ is motivated by the *ab-initio* calculations[23], reflecting the fact that there are minor/no lattice-reconstruction effects at 3°.

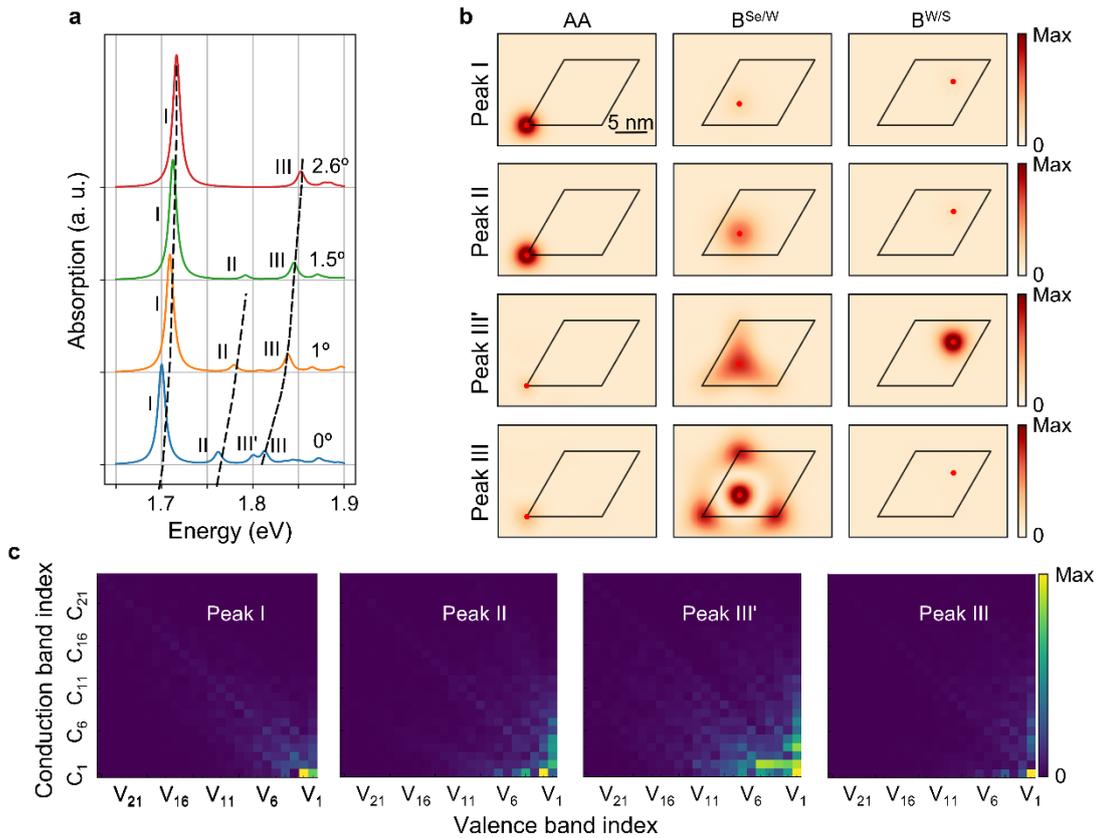

**Fig. S10: (a)** Twist-angle dependence of absorption spectrum obtained from continuum-model wavefunctions; **(b)** for the 0° twist angle, electron-density distributions for a fixed hole position, for all excitonic peaks, where peak III′ denotes the peak with energy slightly lower than peak III; and **(c)** corresponding k-space–averaged distributions ($\frac{1}{N_k}\sum_k |A^s_{vck}|$) of each exciton(s) across the valence and conduction bands.

Fig. S10 shows that at twist angle close to 0°, the exciton energies and wavefunctions calculated using this model qualitatively match the *ab initio* results presented in the main text Fig. 2. For twist angles slightly higher than 0°, we find that the overall character of excitonic wavefunctions remains unchanged (see Fig. S11).

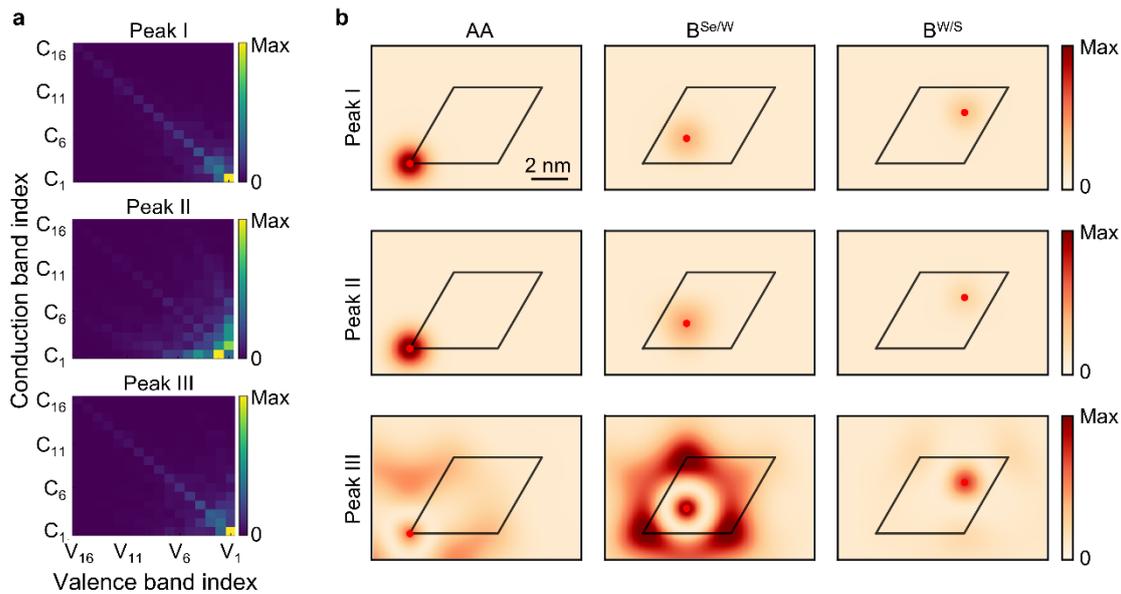

**Fig. S11: (a)** k-space–averaged valence and conduction-band distributions of each exciton at 1.45° twist angle, and **(b)** the corresponding electron-density distributions for fixed hole positions.

**Section 10: Correlation between peak I and biexciton**

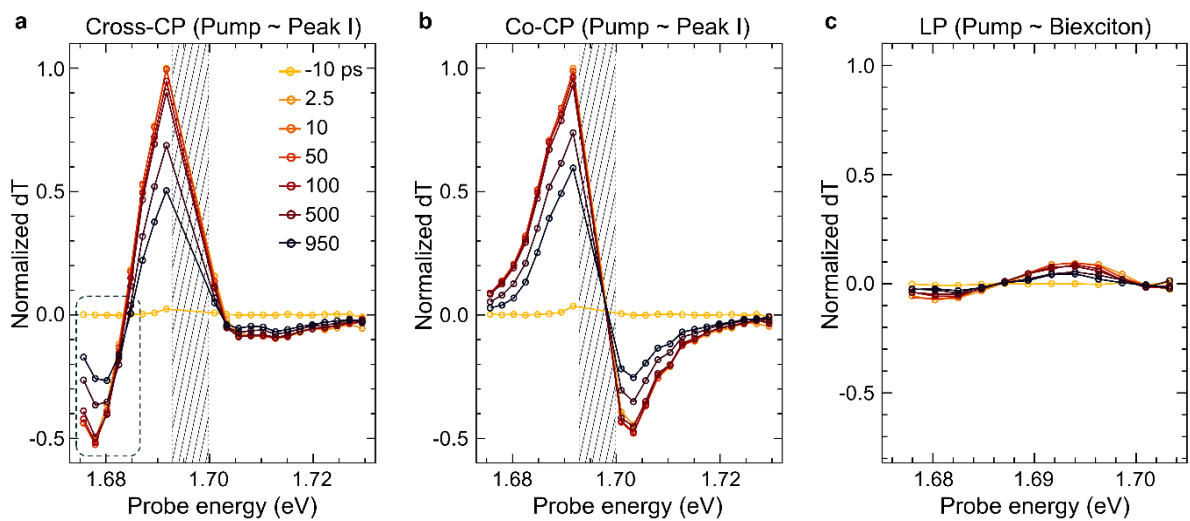

**Fig. S12:** Based on the cross- and co-circular polarization (CP) maps, **(a)** and **(b)** show linecuts along the pump energy resonating with peak I at various delay times. A strong, low-energy state at 1.678eV is clearly visible

(shown in dotted box) in the cross-CP map but is absent in the co-CP map (the negative signal at higher energy is a pump-induced energy shift as discussed in Section 12). Note that the data in shaded area is inaccessible because the spectrometer is used to block the pump before the detector. **(c)** displays linecuts taken from linearly polarized (LP) maps along the pump in resonance with biexciton. At this energy, both the biexciton and peak I signals are very weak, which suggests that peak I is responsible for the biexciton state, directly demonstrating their correlation.

**Section 11: Dynamics of biexcitonic state**

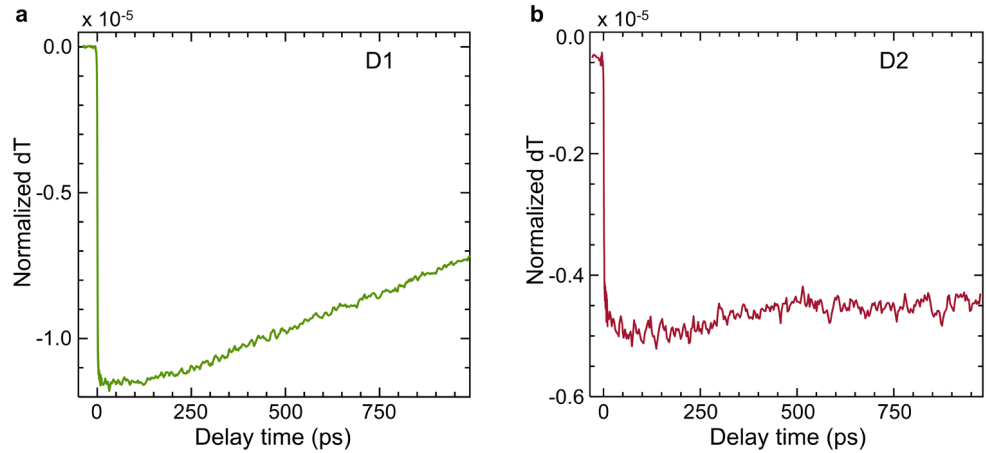

**Fig. S13:** **(a)** and **(b)** show the long-lived biexcitonic state observed in cross-CP pump-probe measurement with pumping at peak I resonance and probing at 1.678 eV. The binding energy for biexcitons is ~ 16 meV in both D1 and D2.

**Section 12: Peak shifts in TCPP maps**

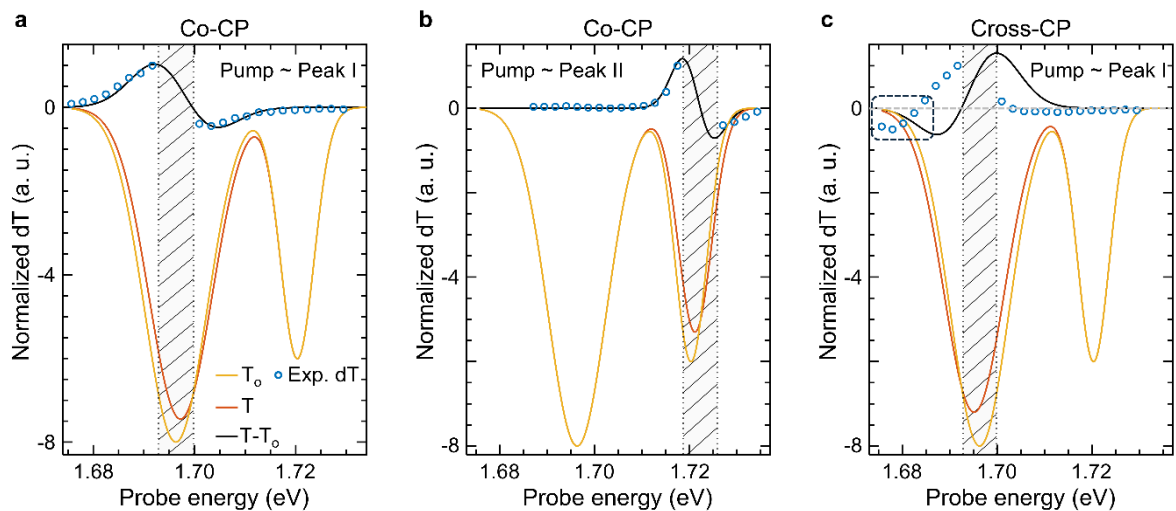

**Fig. S14:** **(a)** and **(b)** show the linecuts from co-CP map along pump energies in resonance with peak I and II, respectively. $T_o$ is the transmission when the pump is off and T is the transmission when the pump is on. When peaks I and II are slightly blue-shifted and broadened in T, it fits precisely with the experimental data. This blue shift can be caused by the Burstein-Moss effect-induced Pauli blocking[2,29,30]. The pump-induced absorption (PA)

signals towards higher energies than peaks I and II are pump-induced energy shifts, as these can be fitted well with T-T$_o$. In contrast, **(c)** shows the linecut from the cross-CP map along pump energy resonant to peak I, highlights the new PA feature at energies lower than Peak I (in dotted box). This is interpreted as a biexcitonic state, since it cannot be fitted by any peak shifts and full width at half maximum (FWHM) changes in T.

For fitting, T and T$_o$ are considered as the sum of two Gaussians corresponding to peaks I and II. T$_o$ is fixed with peaks at 1696 meV and 1720 meV, having FWHM of 8.7 meV and 4.7 meV, respectively. For Fig. S14a, T has a slightly blue-shifted peak I at 1697.2 meV with FWHM 8.9 meV; for Fig. S14b, the blue-shifted peak II is at 1721.2 meV with FWHM 5.0 meV, which indicates that energy shifts are contributing more.

**Section 13: Raw data and decay subtraction of peak II to extract oscillations**

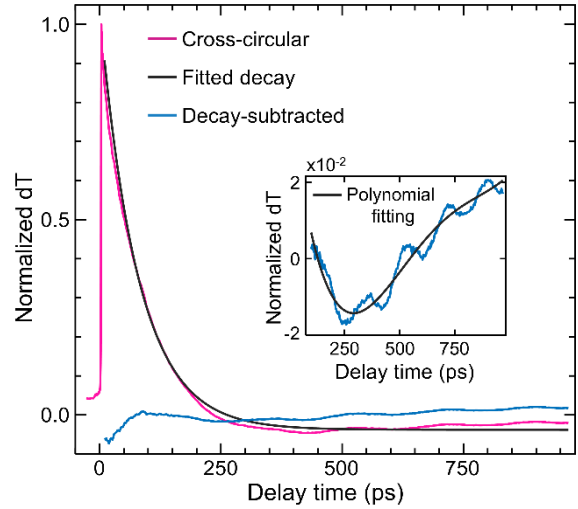

**Fig. S15:** Raw TCPP data for device D1 at the resonance of peak II, using cross-CP. An exponential decay is fitted and then subtracted from the raw data to extract the oscillations. To calculate FFT, a polynomial background is subtracted from the extracted oscillations (inset).

**Section 14: Identification of GHz oscillations as phasons**

Here, we explore other possible explanations for dT oscillations. Magnons, the collective spin excitations, would exhibit a strong dependence on circular polarization due to their spin-dependent nature[31], and are not intrinsically present in non-magnetic TMDs heterostructures. Plasmons, collective oscillations of charge carriers, typically have much higher energies (in THz to mid-infrared range)[32], and are usually relevant in the proximity of metal nanoparticles. Polaritons, exciton-photon quasiparticles, are expected to have very high energy[33]. Hence, all these modes are ruled out from our observations.

Conversely, phasons, the collective, low-energy modes associated with the relative sliding motion of the moiré superlattice, align with our experimental results. Their frequency is a direct function of the twist angle, and their nature as a lattice translation makes them insensitive to the spin-valley degree of freedom and consequently, to the circular polarization of the excitation light.

**Section 15: Phason modes for hBN encapsulated WSe$_2$/WS$_2$**

As discussed in the main text, we performed phonon calculations at 0° twist angle and identified shear-like phonon modes with energies below 1 cm$^{-1}$, which we attribute to phason modes[34,35]. The corresponding phason-mode eigenvectors are strongly localized in AA-stacked regions.

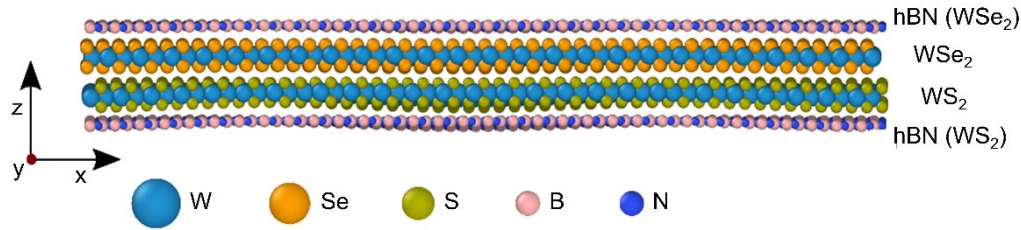

**Fig. S16:** Figure showing structure of hBN encapsulated WSe$_2$/WS$_2$.

We further examined the effect of hBN encapsulation on these modes, since the presence of a substrate could, in principle, modify phason localization. In the hBN-encapsulated structure (Fig. S16), where the WS$_2$/WSe$_2$ heterostructure is sandwiched between hBN layers on each side, we find that out-of-plane relaxation is substantially suppressed, whereas the in-plane components of the phason eigenvectors, as well as their real-space localization pattern, remain essentially unchanged. This indicates that the phason-mode localization is robust and persists even in the presence of hBN encapsulation (or a substrate). In this case, there are nine low-energy modes with energies below 1 cm$^{-1}$: three acoustic modes of whole heterostructure (in x, y and z direction) and six phason modes corresponding to relative in-plane shifts of the different layers, consistent with the expectation that each layer contributes two in-plane degrees of freedom, giving a total of eight in-plane modes, of which two are acoustic and six are shear-like. Among these, the four modes primarily associated with shifting of hBN layers have energies very close to zero. As shown in Fig. S17, the phason modes with dominant hBN character are relatively delocalized, and therefore not expected to significantly affect the excitonic states or optical absorption, and thus not lead to the oscillations observed in pump-probe spectroscopy. As shown in Fig. S17, two modes exhibit slightly negative energies of approximately -1 cm$^{-1}$, which arise from numerical artifacts and are additionally influenced by the accuracy of the potentials used in computing the force constants.

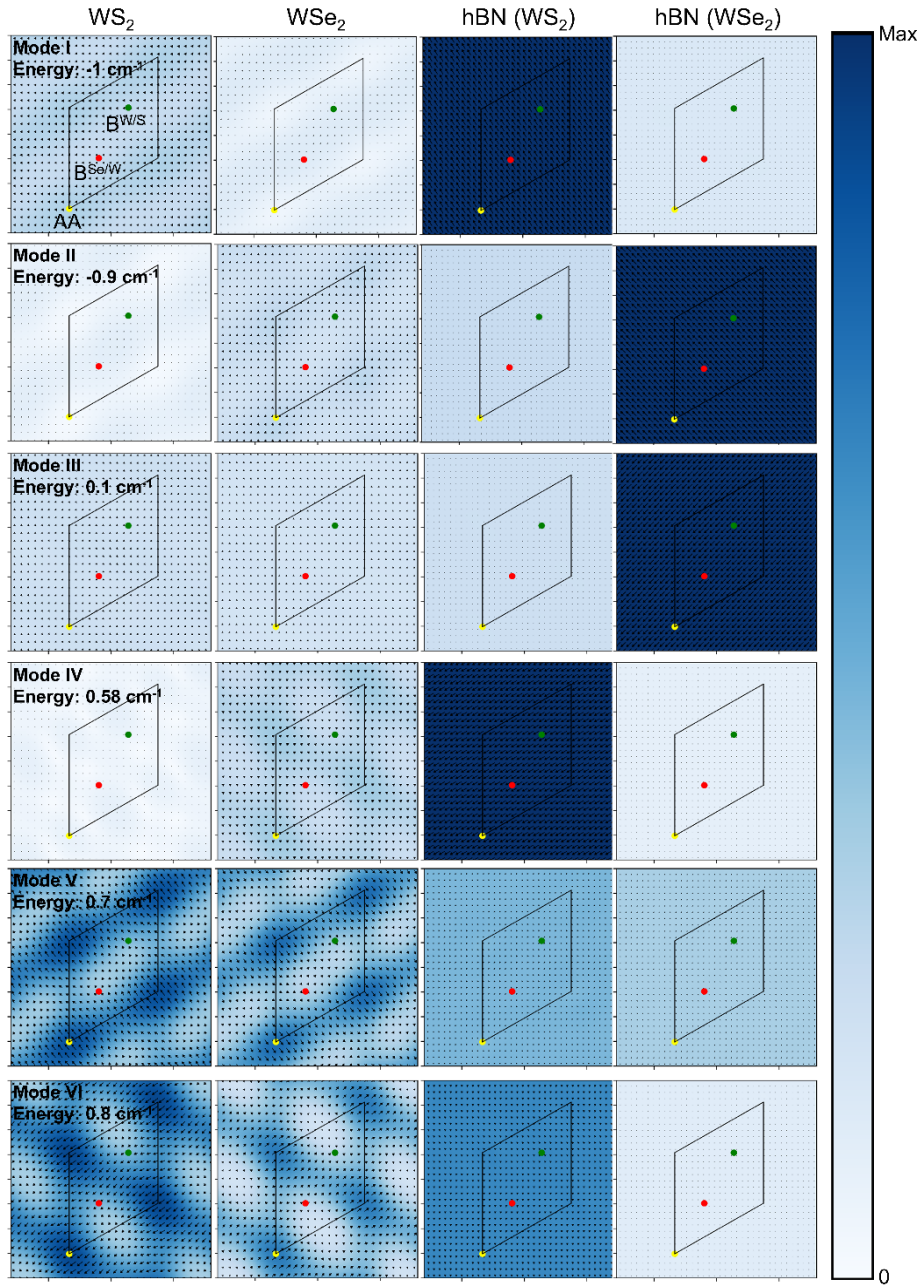

**Fig. S17:** In-plane phason eigenvectors for hBN-encapsulated WSe$_2$/WS$_2$, where arrows indicate the in-plane eigenvector direction, and their length and colour both represent its magnitude; each column corresponds to the eigenvector components of a different layer, with hBN (WSe$_2$/WS$_2$) denoting the hBN layer placed adjacent to either the WSe$_2$ or the WS$_2$ layer. For clarity, arrows are shown only for W atoms in the WSe$_2$/WS$_2$ layers and for B atoms in the hBN layer, and the arrow lengths in hBN are reduced by a factor of two relative to those in WSe$_2$/WS$_2$ to improve visual clarity.

We investigated the twist-angle dependence of phason-mode energies by performing phonon calculations at three twist angles: 0°, 1.43°, and 2.6°. The results are shown in Fig. 4d of the main text. For this study, we employed a hexagonal WS$_2$/WSe$_2$ bilayer unit cell without hBN encapsulation. The atomic structures were generated using the *Twister* package[6] and subsequently relaxed using LAMMPS[10] until the residual forces on all atoms were reduced below $10^{-7}$ eV Å$^{-1}$ in each Cartesian direction (compared to $10^{-6}$ eV Å$^{-1}$ for the hBN-encapsulated structure). Force-constant generation and phonon calculations were carried out with PARPHOM[36]. In this configuration, we identified only two phason modes, which are nearly degenerate and correspond to the relative in-plane shear-like motion at the single WS$_2$/WSe$_2$ interface.

**Section 16: GHz oscillations in other states**

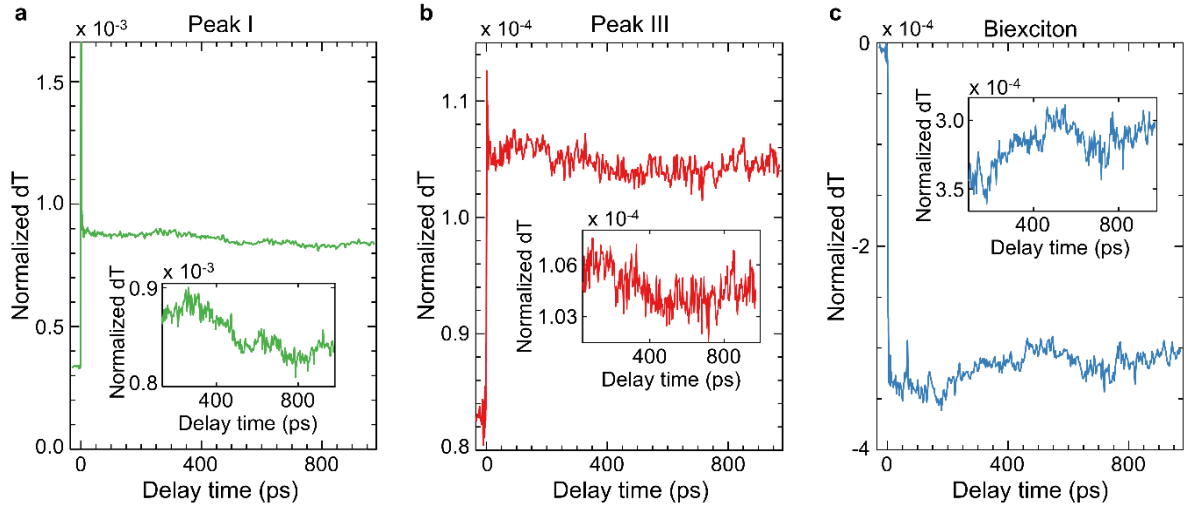

**Fig. S18: (a)**, **(b)** and **(c)** show the weak oscillations in peak I, peak III, and the biexciton state, respectively, and the insets show their zoomed-in regions.